\documentclass[twocolumn,pre,amsmath,amssymb,superscriptaddress]{revtex4}
\usepackage[pdftex]{graphicx,color}
\usepackage{color}
\usepackage{float}
\usepackage{verbatim}
\usepackage{stmaryrd}
\usepackage{calc}
\usepackage{epsfig} 
\usepackage{subfigure} 
\usepackage{bm}% bold math
\def \equi#1{\mathrel{\mathop{\kern 0pt\sim}\limits_{#1}}} 
\scrollmode

\begin{document}
\title{Search in Patchy Media: Exploitation-Exploration Tradeoff}

\author{M. Chupeau} 
\affiliation{Laboratoire de Physique Th\'eorique et Mod\`eles Statistiques (UMR 8626), Universit\'e Paris-Sud, Universit\'e Paris-Saclay, 91405 Orsay, France}
\affiliation{Laboratoire de Physique Th\'eorique de la
  Mati\`ere Condens\'ee (UMR 7600), Universit\'e Pierre et Marie Curie,
  4 Place Jussieu, 75255 Paris Cedex France}

\author{O. B\'enichou}
\affiliation{Laboratoire de Physique Th\'eorique de la
  Mati\`ere Condens\'ee (UMR 7600), Universit\'e Pierre et Marie Curie,
  4 Place Jussieu, 75255 Paris Cedex France}

\author{S. Redner} \affiliation{Santa Fe Institute, 1399 Hyde Park Road, Santa Fe, New
  Mexico 87501, USA and Department of Physics, Boston University, Boston,
  MA 02215, USA}
  
\begin{abstract}

  How to best exploit patchy resources? We introduce a minimal
  exploitation/migration model that incorporates the coupling between a
  searcher's trajectory, modeled by a random walk, and ensuing depletion of
  the environment by the searcher's consumption of resources.  The searcher
  also migrates to a new patch when it takes $\mathcal{S}$ consecutive steps
  without finding resources.  We compute the distribution of consumed
  resources $F_t$ at time $t$ for this non-Markovian searcher and show that
  consumption is maximized by exploring multiple patches.  In one dimension,
  we derive the optimal strategy to maximize $F_t$.  This strategy is robust
  with respect to the distribution of resources within patches and the
  criterion for leaving the current patch.  We also show that $F_t$ has an
  optimum in the ecologically-relevant case of two-dimensional patchy
  environments.
  
\end{abstract}
%\pacs{05.40.Jc, 87.23.Cc}
\maketitle

\section{Introduction}

\begin{figure}[!h]
\centering
\includegraphics[width=240 pt]{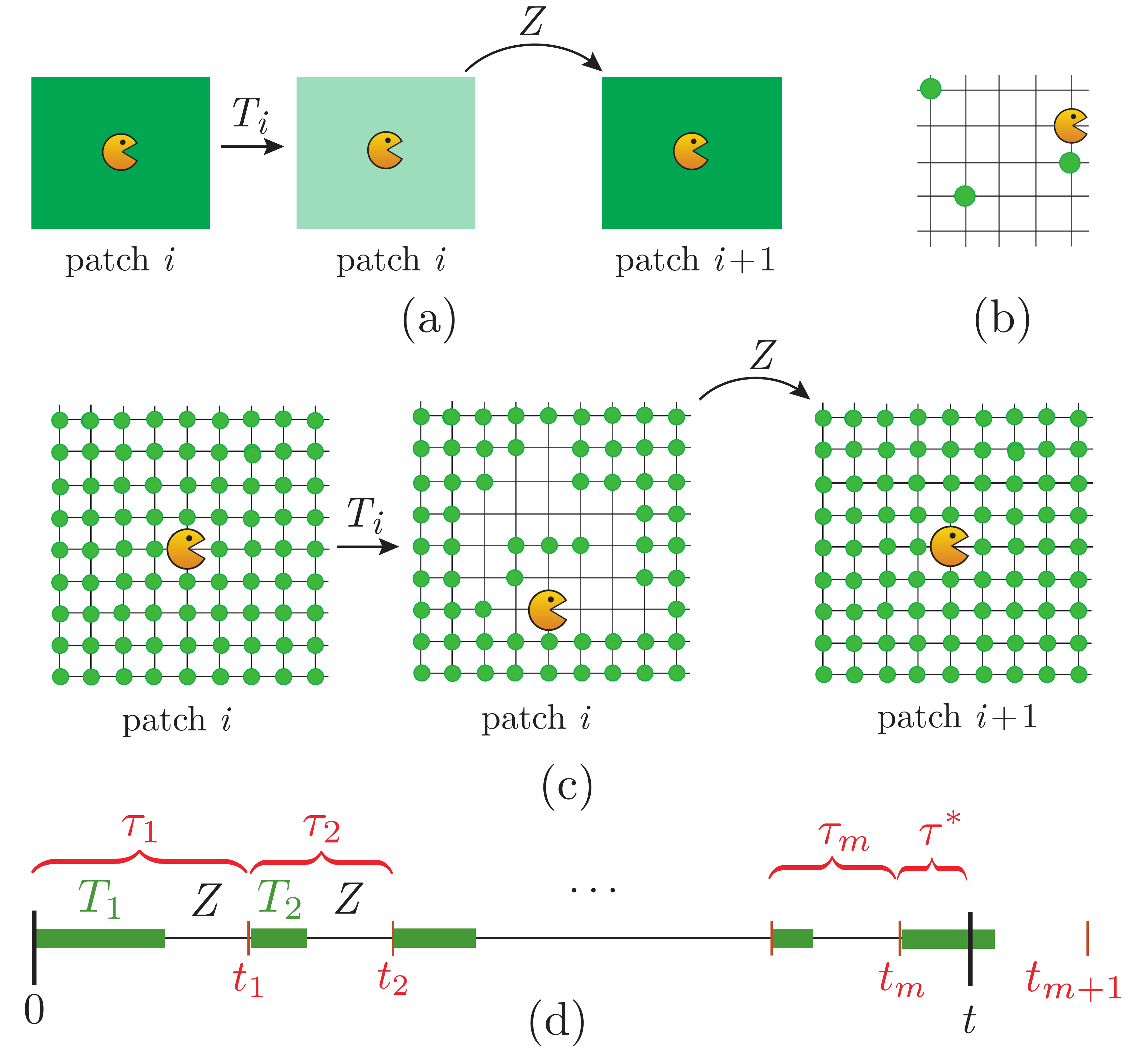}
\caption{(a) Continuous patch use: a searcher uniformly depletes patch $i$ at
  a fixed rate for a deterministic time $T$ and moves to a patch $i+1$ when
  patch $i$ is sufficiently depleted.  (b) Random search: a searcher seeks
  one or a few fixed targets (circles) via a random walk.  (c) Our model: a
  searcher depletes resources within a patch for a random time $T_i$. (d)
  Model time history.  Phase $i$, of duration $\tau_i$, is composed of patch
  exploitation (duration $T_i$, shadowed) and migration (duration $Z$). The
  last phase is interrupted at time $t$, either during exploitation (shown
  here) or migration, and lasts $\tau^*$.}
\label{model}
\end{figure}

Optimizing the exploitation of patchy resources is a long-standing dilemma in
a variety of search problems, including robotic
exploration~\cite{Wawerla:2009}, human decision
processes~\cite{Gittins:1973}, and especially in animal
foraging~\cite{MP66,Pyke:1977,VLRS11,Charnov:1976}.  In foraging, continuous
patch-use~\cite{Charnov:1976,Stephens-DW:1986} and random
search~\cite{Benichou:2011,VLRS11} represent two paradigmatic exploitation
mechanisms.  In the former (Fig.~\ref{model}(a)), a forager consumes
resources within a patch until a specified depletion level, and concomitant
decrease in resource intake rate, is reached before the forager moves to
another virgin patch.  

In his pioneering work~\cite{Charnov:1976}, Charnov predicted the optimal
strategy to maximize resource consumption.  This approach specifies how
fitness-maximizing foragers should use environmental information to determine
how completely a food patch should be exploited before moving to new foraging
territory.  The nature of foraging in an environment with resources that are
distributed in patches has been the focus of considerable research in the
ecology literature (see e.g.,
\cite{Charnov:1976,Oaten:1977,G80,Iwasa:1981,M82,OH88,MCS89,VB89,V91,OB06,V06,ECMG07,PS11});
theoretical developments are relatively mature and many empirical
verifications of the theory have been found.  However, continuous patch use
models typically do not account for the motion of the searcher within a
patch, and the food intake rate within a patch is given {\it a priori}
\cite{Oaten:1977,Iwasa:1981,Green:1984}, so that depletion is deterministic
and spatially homogeneous.

Random search represents a complementary perspective in which the searcher
typically moves by a simple or a generalized random walk.  The search
efficiency is quantified by the time to reach targets (Fig.~\ref{model}(b)).
Various algorithms, including L\'evy strategies~\cite{Viswanathan:1999a},
intermittent
strategies~\cite{Benichou:2005,Oshanin:2007,Lomholt:2008,Bressloff:2011} and
persistent random walks~\cite{Tejedor:2012}, have been shown to minimize this
search time under general conditions.  However, these models do not consider
depletion of the targets.

Issues that have been addressed to some extent in the above scenarios include
the overall influence of resource patchiness (but
see~\cite{M82,MCS89,ECMG07,PS11,WRVL15,NWS15} for relevant work), as well as
the coupling between searcher motion within patches and resource depletion;
the latter is discussed in a different context than that given here in
Ref.~\cite{RBL03}.  In this work, we introduce a minimal patch
exploitation/inter-patch migration model that accounts for the interplay
between mobility and depletion from which we are able to explicitly derive
the amount of consumed food $F_t$ up to time $t$, determine the optimal
search strategy, and test its robustness.

\section{The Model}

Each patch is modeled as an infinite lattice, with each site initially
containing one unit of resource, or food.  A searcher undergoes a
discrete-time random walk within a patch and food at a site is completely
consumed whenever the site is first visited.  The searcher thus sporadically
but methodically depletes the resource landscape.

Resources within a patch become scarcer and eventually it becomes
advantageous for the searcher to move to a new virgin patch.  We implement
the scarcity criterion that the searcher leaves its current patch upon
wandering for a time $\mathcal{S}$ without encountering food.  Throughout
this work, all times are rescaled by the (fixed) duration of a random-walk
step.  Thus $\mathcal{S}$ also represents the number of random-walk steps
that the walker can take without finding food.  This notion of a specified
``give-up time'' has been validated by many ecological
observations~\cite{Krebs:1974,Iwasa:1981,McNair:1983,Green:1984}.  The
searcher therefore spends a random time $T_i$ and consumes $f_i$ food units
in patch $i$, before leaving (Fig.~\ref{model}(d)).  We assume, for
simplicity, a deterministic migration time $Z$ to go from one patch to the
next.  We define $t_i$ as the time when the searcher arrives at patch $i+1$
and $\tau_i=t_i-t_{i-1}$ as the time interval between successive patch
visits.  The duration of phase $i$, which starts at $t_{i-1}$ and consists of
exploitation in patch $i$ and migration to patch $i+1$, is
$\tau_i \equiv T_i+Z$.

Our model belongs to a class of composite search strategies that incorporate:
(i) intensive search (patch exploitation) and (ii) fast displacement
(migration)~\cite{Benhamou:2007,Plank:2008,Nolting:2015}; here we extend
these approaches to account for resource depletion.  In addition to its
ecological relevance, this exploit/explore duality underlies a wide range of
phenomena, such as portfolio optimization in finance~\cite{Gueudre:2014},
knowledge management and transfer~\cite{March:1991}, research and development
strategies~\cite{Gittins:1973}, and also everyday life decision
making~\cite{Cohen:2007}.

We quantify the exploitation efficiency by the amount of consumed food $F_t$
up to time $t$.  Note that $F_t$ is also the number of distinct sites that
the searcher visits by time $t$, which is known for Markovian random
walks~\cite{Weiss,Hughes}.  In our model, we need to track \emph{all}
previously visited sites in the current patch to implement the scarcity
criterion, which renders the dynamics non-Markovian.

We first argue that $F_t$ admits a non-trivial optimization in spatial
dimensions $d\leq 2$.  If a random-walk searcher remains in a single patch
forever (pure exploitation; equivalently, $\mathcal{S}\to \infty$), then
$F_t$, which coincides with the number of distinct sites visited in the
patch, grows sublinearly in time, as $\sqrt{t}$ in $d=1$ and as $t/\ln t$ in
$d=2$~\cite{Weiss}.  On the other hand, if the searcher leaves a patch as
soon as it fails to find food (pure exploration, $\mathcal{S}=1$), $F_t$
clearly grows linearly in time, albeit with a small amplitude that scales as
$1/Z$.  Thus $F_t$ must be optimized at some intermediate value of
$\mathcal{S}$, leading to substantial exploitation of the current patch
before migration occurs.

\section{The Amount of Food Consumed}

\subsection{Formalism}
\label{formal}

To compute the amount of food consumed, let $m$ be the (random) number of
phases completed by time $t$, while the $(m+1)^{\rm st}$ phase is interrupted
at time $t$.  Then $F_t$ can be written as
\begin{subequations}
\begin{equation}
\label{Ct}
F_t=f_1+\ldots+f_{m}+f^*,
\end{equation}
where $f^*$ denotes the food consumed in this last incomplete phase.
Similarly, the phase durations $\{\tau_i\}$ satisfy the sum rule
(Fig.~\ref{model}(d))
\begin{equation}
\label{sumrule}
t=\tau_1+\ldots+\tau_{m}+\tau^*,
\end{equation}
\end{subequations}
where again $\tau^*$ denotes the duration of the last phase.  Since the food
consumed and the duration of the $i^{\rm th}$ phase, $f_i$ and $\tau_i$
respectively, are correlated, the sum rule~\eqref{sumrule} couples the
$f_i$'s and the number $m$ of patches visited.  The distinct variables $f_i$
and $\tau_i$ are correlated and pairwise identically distributed, except for
the last pair $(f^*,\tau^*)$ for the incomplete phase.  We will ignore this
last pair in evaluating $F_t$, and approximation is increasingly accurate for
large $\mathcal{S}$.  

We now express the distribution of $F_t$ in terms of the joint distribution
of the food consumed in any phase and the duration of any phase, which we
compute in $d=1$.  For this purpose, we extend the approach developed
in~\cite{Godreche:2001} for standard renewal processes to our situation where
$f$ and $\tau$ are coupled.  To obtain of $F_t$, it is convenient to work
with the generating function $\langle e^{-p F_t} \rangle$, where the angle
brackets denote the average over all possible searcher trajectories.  This
includes integrating over each phase duration, as well as summing over the
number of phases and the food consumed in each patch.  The generating
function can therefore be written as
\begin{align}
\label{basis}
\left\langle e^{-pF_t}\right\rangle=&\sum_{m=0}^\infty\int_{\mathbb{R}^{m}} \kern-0.6em {\rm d}y_1\ldots{\rm d}y_m \sum_{n_1,\ldots,n_m} e^{-p(n_1+\ldots.+n_m)}\nonumber\\
&\times {\rm Pr}\big(\{n_i\},\{y_i\},m\big),
\end{align}
where we now treat the time as a continuous variable in the long-time limit.
The second line is the joint probability that the food consumed in each patch
is $\{ n_i\}$, that each phase duration is $\{ y_i\}$, and that $m$ phases
have occurred; we also ignore the last incomplete phase.  From
Fig.~\ref{model}(d), the final time $t$ occurs sometime during the
$(m+1)^{\rm st}$ phase, so that $t_m<t<t_{m+1}$.

We rewrite the joint probability as the ensemble average of the following
expression that equals 1 when the process contains exactly $m$ complete
phases of durations $\{ y_i\}$, with $n_i$ units of food consumed in the
$i^{\rm th}$ phase, and equals 0 otherwise:
\begin{align}
{\rm Pr}\big(\{n_i\},\{y_i\},m\big)%\nonumber\\
\!=\!\Big\langle\! \prod_{i=1}^m\delta_{f_i,n_i}\delta(\tau_i-y_i)I(t_m\!<\!t\!<\!t_{m+1})\Big\rangle,
\end{align}
with the indicator function $I(z)=1$ if the logical variable $z$ is true, and
$I(z)=0$ otherwise.  We can compute the Laplace transform with respect to the
time $t$ of this joint probability (see Appendix~\ref{eq4}) from which the temporal
Laplace transform of the generating function $\langle e^{-p F_t} \rangle$ is
\begin{align}
\label{step1}
\int_0^\infty {\rm d}t \, e^{-st}\left\langle e^{-pF_t}\right\rangle 
=\frac{1-\left\langle e^{-s\tau} \right\rangle_1}
{s\left(1-\left\langle e^{-pf-s\tau}\right\rangle_1 \right)}.
\end{align}
Here $\left\langle e^{-pf-s\tau}\right\rangle_1$ is an ensemble average over
the values $(f,\tau)$ for the amount of food consumed in a \emph{single}
phase and the duration of this phase; we use the subscript 1 to indicate such
an average over a single phase.  Equation~\eqref{step1} applies for any
distribution of the pair $(f,\tau)$; in particular for any spatial dimension,
search process, and distribution of food within patches.

\subsection{Detailed Results}

We now make Eq.~\eqref{step1} explicit in $d=1$ by calculating
$\left\langle e^{-s\tau-pf}\right\rangle_1$.  For this purpose, we make use
of the equivalence between exploitation of a single patch and the survival of
a starving random walk~\cite{Benichou:2014,CBR16}.  In this latter model, a
random walk is endowed with a metabolic capacity $\mathcal{S}$, defined as
the number of steps the walker can take without encountering food before
starving.  The walker moves on an infinite $d$-dimensional lattice, with one
unit of food initially at each site.  Upon encountering a food-containing
site, the walker instantaneously and completely consumes the food and can
again travel $\mathcal{S}$ additional steps without eating before starving.
Upon encountering an empty site, the walker comes one time unit closer to
starvation. 

In our exploitation/migration model, the statistics of $(f,\tau)$ for a
searcher that leaves its current patch after $\mathcal{S}$ steps coincides
with the known number of distinct sites visited and lifetime of a starving
random walk with metabolic capacity $\mathcal{S}$ at the instant of
starvation~\cite{Benichou:2014,CBR16}.  In Appendix~\ref{eq5ab}, we determine the
full distribution of the pair $(f,\tau)$, from which we finally extract the
quantity $\left\langle e^{-pf-s\tau}\right\rangle_1$ in Eq.~\eqref{step1},
where $\tau=T+Z$, with $T$ the (random) time spent in a patch and $Z$ the
fixed migration time.  The final result is
\begin{subequations}
\begin{align}
\label{step2}
\langle e^{-pf-s\tau}\rangle_1=\int_0^\infty \!\! {\rm d}\theta  
\,P(\theta) \,\,e^{\,[-p\pi\theta\sqrt{\mathcal{S}/2}-s(Z+\mathcal{S})+Q(\theta)]}\,,
\end{align}
where
\begin{align}
\begin{split}
\label{Q}
Q(\theta)&=\exp\bigg[4\int_0^\theta\frac{{\rm d}u}{u}\sum_{j=0}^\infty q_j\bigg]\,,\\
q_j&=\frac{1-e^{-[s\mathcal{S}+{(2j+1)^2}/{u^2}]}}{1+{su^2\mathcal{S}}/{(2j\!+\!1)^2}}
-\left(1-e^{-{(2j+1)^2}/{u^2}}\right)\,,\\
P(\theta)&=\frac{4}{\theta}\sum_{j= 0}^\infty e^{-(2j+1)^2/\theta^2}
\exp\!\bigg[\!-2\!\sum_{k= 0}^\infty E_1
  \big({(2k+1)^2}/{\theta^2}\big)\bigg],
\end{split}
\end{align}
\end{subequations}
and $E_1(x)=\int_1^\infty{\rm d}t \, e^{-xt}/t$ is the exponential integral.

\begin{figure}[h!]
\centering
\includegraphics[width=250 pt]{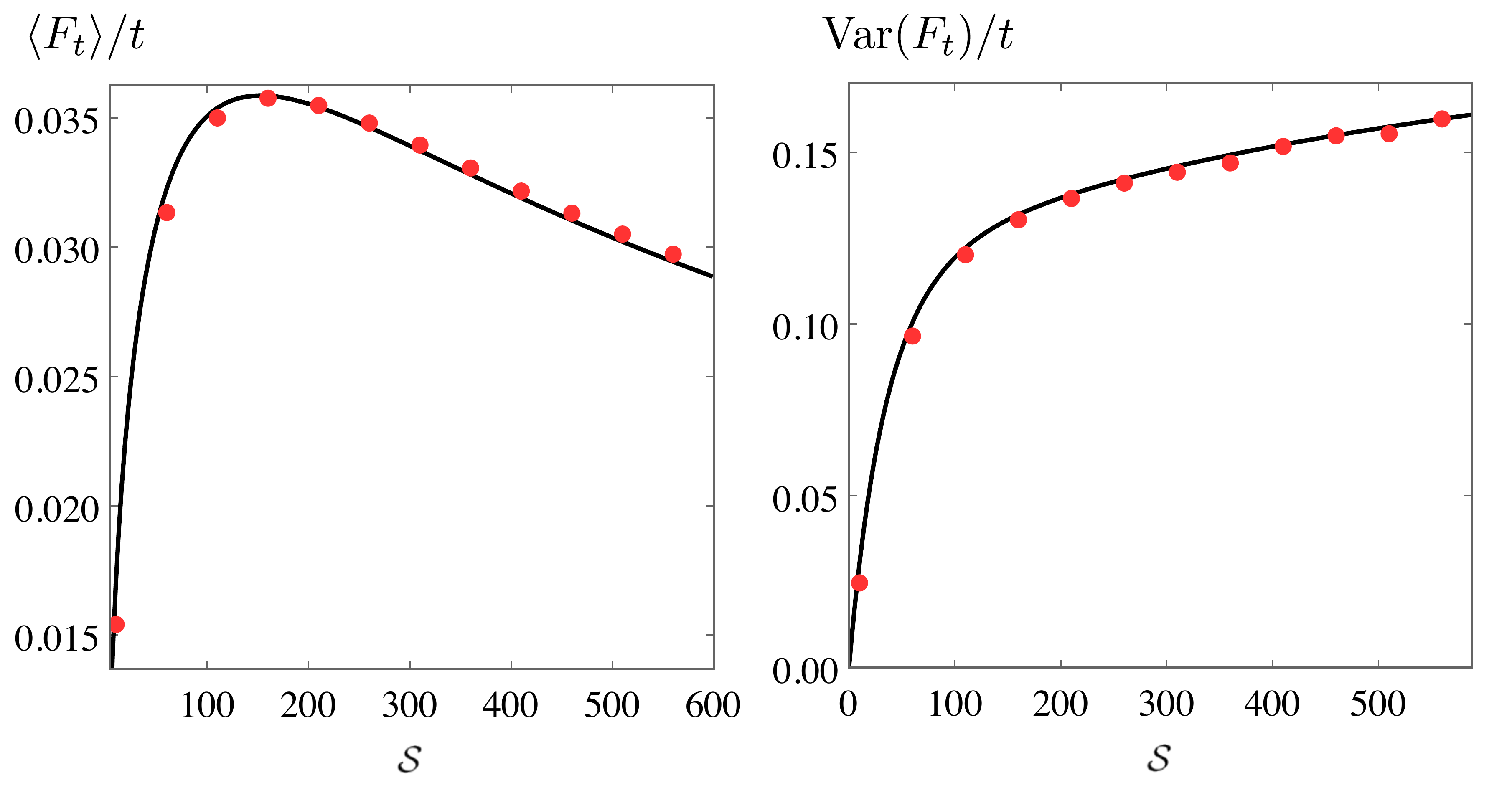}
\caption{Scaled mean (a) and variance (b) of the food consumed $F_t$ at
  $t=5\times 10^5$ steps.  Points give numerical results and the curves are
  the asymptotic predictions in \eqref{mean-var2}.  The migration time $Z$
  between patches is 500 steps.}
\label{moments}
\end{figure}

We now focus on the first two moments of $F_t$, whose Laplace transforms are
obtained from the small-$p$ expansion of Eq.~\eqref{step1}.  By analyzing
this expansion in the small-$s$ limit, the long-time behavior of these
moments are (with all details given in Appendix~\ref{eq6}):
\begin{align}
\begin{split}
\label{mean-var1}
&\frac{\langle F_t\rangle}{t} \sim\frac{\langle f\rangle}{\langle T\rangle +Z}\,, \\[1em]
&\frac{{\rm Var} (F_t)}{t} \!\sim\!\frac{\langle f\rangle ^2{\rm
    Var}(T)}{(\langle T\rangle \!+\!Z)^3}+ \frac{{\rm Var}(f)}{\langle T\rangle \!+\!Z}
-2\frac{\langle f\rangle {\rm Cov}(f,T)}{(\langle T\rangle \!+\!Z)^2}\,,
\end{split}
\end{align}
where ${\rm Var}(X)\equiv\langle X^2\rangle -\langle X\rangle^2$ and
${\rm Cov}(X,Y)\equiv\langle XY\rangle-\langle X\rangle\langle Y\rangle$ and
for simplicity, we now drop the subscript 1.  From the small-$p$ and
small-$s$ limits of Eqs.~\eqref{step2} and~\eqref{Q}, the limiting behavior
of the moments for $\mathcal{S}\gg 1$ are:
\begin{align}
\begin{split}
\label{mean-var2}
&\frac{\langle F_t\rangle}{t} \simeq\frac{K_1\sqrt{\mathcal{S}}}{K_2\mathcal{S}+Z}\,,  \\[0.1in]
&\frac{{\rm Var} (F_t)}{t} \simeq \left[\frac{K_3 \mathcal{S}^3}{(K_2\mathcal{S}\!+\!Z)^3}
+\frac{K_4 \mathcal{S}}{K_2\mathcal{S}\!+\!Z}
-\frac{K_5 \mathcal{S}^2}{(K_2\mathcal{S}\!+\!Z)^2}\right]\,,
\end{split}
\end{align}
where the $K_i$ are constants that are derived in Appendix~\ref{eq7}.  The dependences
$\langle f\rangle=K_1\sqrt{\mathcal{S}}$ and
$\langle T\rangle=K_2\mathcal{S}$ have simple heuristic explanations (see
also~\cite{CBR16}): suppose that the length of the interval where resources
have been consumed reaches a length $\sqrt{\mathcal{S}}$.  When this critical
level of consumption is reached, the forager will typically migrate to a new
patch because the time to traverse the resource-free interval will be of the
order of $\mathcal{S}$.  Thus the resources consumed in the current patch
will be of the order of the length of the resource-free region, namely
$\sqrt{\mathcal{S}}$, while the time $\langle T\rangle$ spent in this patch
will be of the order of the time $\mathcal{S}$ to traverse this region of
length $\sqrt{\mathcal{S}}$.

The salient feature from Eq.~\eqref{mean-var2} is that $\langle F_t\rangle$
has a maximum, which occurs when $\mathcal{S}=Z/K_2$, corresponding to
$\langle T \rangle = Z$ (Fig.~\ref{moments}). That is, the optimal strategy
to maximize food consumption is to spend the same time exploiting each patch
and migrating between patches.  
%At optimality, Eqs.~\eqref{mean-var2} also
%give, for the normalized variance,
%\begin{equation}
%\frac{{\rm Var}(F_t)}{\langle F_t \rangle^2} \simeq \alpha \frac{Z}{t}\,,
%\end{equation}
%with $\alpha \approx 0.203$ to three significant figures.  
% Thus relative fluctuations between individual trajectories are not small and
% increase with $Z$.  That is, the amount of food consumed in a patchy
% environment is a highly variable process.

It is worth mentioning that we can reproduce the first of
Eqs.~\eqref{mean-var1} by neglecting correlations between $f$ and $T$.  In
this case $\langle F_t\rangle$ is simply the average amount of food
$\langle f \rangle$ consumed in a single patch multiplied by the mean
number $t/(\langle T\rangle+Z)$ of patches explored at large time $t$.
However, this simple calculational approach fails to account for the role of
fluctuations, specifically the covariance between $f$ and $T$, in the
variance of $F_t$.  In fact, the covariance term (last term in
Eq.~\eqref{mean-var1}) reduces fluctuations in food consumption by a factor
three compared to the case where correlations are neglected.

\section{Extensions}

\begin{figure}[h!]
\centering
\includegraphics[width=250 pt]{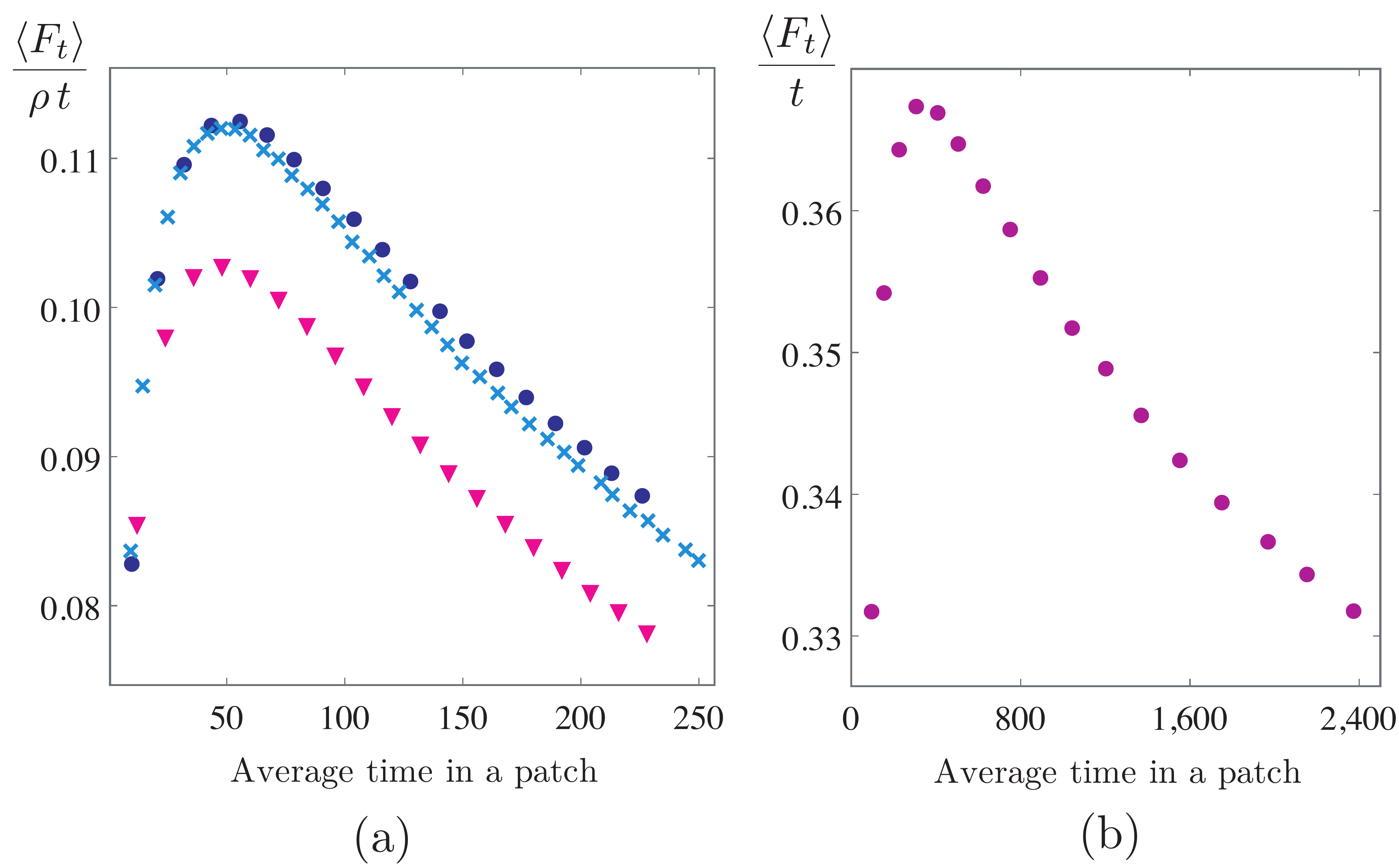}
\caption{Average food consumed in $d=1$ when: (a) the distribution of food is
  Poisson distributed with density $\rho=0.1$ (\textcolor{blue}{$\times$})
  and $\rho=1$ ($\bullet$), as well as when the searcher has a constant
  probability at each step to leave the patch
  (\textcolor{red}{$\blacktriangledown$}); (b) Average food consumed in $d=2$
  for food density $\rho=1$.  The inter-patch travel time $Z=50$ for all
  cases.}
\label{extensions}
\end{figure}

The optimal strategy outlined above is robust and holds under quite general
conditions, including, for example: (i) randomly distributed food within a
patch, and (ii) searcher volatility.  For (i), suppose that each lattice site
initially contains food with probability $\rho$.  To show that the optimal
search strategy is independent of $\rho$, we again exploit the mapping onto
starving random walks in the limit $\mathcal{S}\gg 1/\rho$.  A density of
food $\rho$ corresponds to an effective lattice spacing that is proportional
to $\rho^{-1/d}$, with $d$ the spatial dimension.  For large $\mathcal{S}$,
this effective lattice spacing has a negligible effect on the statistics of
the starving random walk.  Both the mean lifetime and mean number of distinct
sites visited are the same as in the case where the density of food equals 1.
However, because the probability to find food at a given site is $\rho$, the
amount of food consumed differs from the number of distinct sites visited by
an overall factor $\rho$.  Thus the food consumed at time $t$ (the first of
Eqs.~\eqref{mean-var2}) is simply 
\begin{equation}
\frac{\langle F_t\rangle}{t} \simeq \rho\, \frac{K_1\sqrt{\mathcal{S}}}{K_2\mathcal{S}+Z} \,.
\end{equation}
Consequently, the optimal search strategy occurs for the same conditions as
the case where each site initially contains food (Fig.~\ref{extensions}(a)).

For the second attribute, suppose that the searcher has a fixed probability
$\lambda$ to leave the patch at each step, independent of the current
resource density, rather than migrating after taking $\mathcal{S}$ steps
without encountering food.  The residence time of the searcher on a single
patch thus follows an exponential distribution with mean $\lambda^{-1}$.  The
exploitation of a single patch can now be mapped onto the \emph{evanescent}
random walk model, in which a random walk dies with probability
$\lambda$ at each step~\cite{Yuste:2013}, and for which the mean number of distinct sites
visited has recently been obtained in one dimension.  Since
Eq.~\eqref{step1}, and thus Eqs.~\eqref{mean-var1}, still hold for any
distribution of times spent in each patch, we can merely transcribe the
results of~\cite{Yuste:2013} (in particular their Eq.~(7) and the following
text) to immediately find that the average food consumed at time $t$ is
\begin{align}
\frac{\langle F_t\rangle}{t}\sim\,\,\frac{\sqrt{\coth{{\lambda}/{2}}}}{Z+\lambda^{-1}}~.
\end{align}
Now $\langle F_t \rangle$ is maximized for $1/\lambda\simeq Z$ in the
$Z \gg 1$ limit.  Again, the optimal strategy is to spend the same amount of
time on average in exploiting a patch and in migrating between patches
(Fig.~\ref{extensions}(a)).

For the ecologically-relevant case of two-dimensional resource patches, the
average amount of food consumed is governed by a similar optimization as in
$d=1$ (Fig.~\ref{extensions}(b)).  While the description of the
two-dimensional case does not appear to be analytically tractable, we
numerically find that the optimal strategy consists in spending somewhat more
time exploiting a single patch rather than migrating between patches.  This
inclination arises because patch exploitation---whose efficiency is
quantified by the average number of distinct sites visited by a given
time---is relatively more rewarding in two than in one
dimension~\cite{Weiss,Hughes}.

\section{Summary}

To summarize, we introduced a minimal patch exploitation/inter-patch migration
model that quantifies the couplings between searcher motion within patches,
resource depletion, and migration to new patches.  Our model may provide a
first step to understand more realistic ecological foraging, where effects
such as predation of the forager~\cite{H79,R10}, heterogeneous travel times
between patches~\cite{R12}, and more complex motions than pure random
walks~\cite{RB09,BCRLMC16} are surely relevant.  On the theoretical side, our
model can also be viewed as a resetting process, in which a random walker
stochastically resets to a new position inside a virgin patch.  In contrast
to existing
studies~\cite{Evans:2011a,Evans:2011b,Pal:2015,Nagar:2015,Eule:2016}, the
times between resets are not given {\it a priori} but determined by the walk
itself.  This modification may open a new perspective in the burgeoning area
of resetting processes.

Financial support for this research was provided in part by starting grant
FPTOpt-277998 (OB), by grants DMR-1608211 and DMR-1623243 from the National
Science Foundation and from the John Templeton Foundation (SR), and by the
Investissement d'Avenir LabEx PALM program Grant No.\ ANR-10-LABX-0039-PALM
(MC).

\appendix

\begin{widetext}

\section{Derivation of Eq.~\eqref{step1}}
\label{eq4}

We first provide the details to relate the distribution of the amount of food
$F_t$ consumed by time $t$ to the statistics of the amount of food $f$
consumed in a single patch and the associated phase duration $\tau$ and
thereby obtain Eq.~\eqref{step1}.  Next, in Appendix B, we determine the
joint statistics of the time $T$ spent in a patch and the quantity $f$ of
food consumed in this patch for the case of one-dimensional patches
(Eqs.~\eqref{step2} and \eqref{Q}).  For this calculation, we use the
scarcity criterion that the searcher migrates to a new patch when it takes
$\mathcal{S}$ steps without encountering food.  This criterion maps the
multipatch foraging problem onto the starving random walk model.  In Appendix
C, we extract the moments of $F_t$ from Eq.~\eqref{step1} in the long-time
limit to obtain Eq.~\eqref{mean-var1}.  Combining all these elements
ultimately gives Eqs.~\eqref{mean-var2}, as shown in Appendix D.

% From this expression, we then extract the moments of $F_t$ in the long-time
% limit to obtain Eq.~\eqref{mean-var1} in the main text.  Finally, we
% determine the statistics of $F_t$ in one dimension, by using the scarcity
% criterion that the searcher migrates to a new patch when it takes
% $\mathcal{S}$ steps without encountering food, which maps the multipatch
% foraging problem onto the starving random walk model.  This derivation will
% ultimately give Eqs.~\eqref{step2}--\eqref{mean-var2} in the main text.

We first need to compute the quantity $\langle e^{-pf-s\tau} \rangle_1$ that
appears as Eq.~\eqref{step1}, as well as the moments of $\tau=T+Z$ and $f$
that appear as Eqs.~\eqref{mean-var1}.  Note that all the times in the
problem---$t$, $\mathcal{S}$, $Z$, $T_i$, $\tau_i$ and $t_i$---are rescaled
by the duration of a random-walk step.  We start with the amount of food
$F_t$ consumed by time $t$, which is the sum of the amounts of food consumed
during each of the $m$ complete phases and the last incomplete phase
\begin{equation}
\label{Ft-SI}
F_t=f_1+\ldots+f_m+f^*.
\end{equation}
Since $t$ occurs anytime during the last phase (either during the
exploitation of the $(m+1)^{\rm st}$ patch or during migration to the next
patch), the statistics of the food consumed during the last incomplete phase,
$f^*$, is different from those of all the other $f_i$'s. In the long-time
limit, $f^*$ is negligible compared to the total amount of food consumed
$F_t$, so we ignore $f^*$ in \eqref{Ft-SI} henceforth.

As defined in Eq.~\eqref{basis}, the generating function of $F_t$ is
\begin{equation}
\label{basisSI}
\left\langle e^{-pF_t}\right\rangle=\sum_{m=0}^\infty
\int_{\mathbb{R}^{m}}  {\rm d}y_1\ldots{\rm d}y_m 
\sum_{n_1,\ldots,n_m} e^{-p(n_1+\ldots.+n_m)} \; 
{\rm Pr}\big(\{n_i\},\{y_i\},m\big)\,.
\end{equation}
Here, the joint probability for the amount of food $\{ n_i\}$ consumed in the
$i^{\rm th}$ patch, the times $\{ y_i\}$ for the phase durations, and the
number $m$ of complete phases is
\begin{equation}
\label{jointSI}
{\rm Pr}\big(\{n_i\},\{y_i\},m\big)
\!=\!\Big\langle\! \prod_{i=1}^m\delta_{f_i,n_i}
\delta(\tau_i-y_i)I(t_m\!<\!t\!<\!t_{m+1}\Big\rangle,
\end{equation}
where the indicator function $I(z)=1$ if the logical variable $z$ is true,
and $I(z)=0$ otherwise.  Following~\cite{G01a}, we compute the multiple
Laplace transform of this joint probability with respect to the time $t$ and
the $\{ y_i\}$.  This Laplace transform is
\begin{subequations}
\begin{align}
\label{long}
\mathcal{L}_{t,y_1,...,y_m} {\rm Pr}(\{n_i\},\{y_i\},m)  
\equiv\int_{(\mathbb{R}^+)^{m+1}}{\rm d}t \, {\rm d}y_1...{\rm d}y_m \, 
e^{-(st+u_1y_1+...+u_m y_m)} \; {\rm Pr}\big(\{n_i\},\{y_i\},m\big),
\end{align}
where $s$ is the variable conjugate to $t$, and each $u_i$ is conjugate to
the corresponding $y_i$.  Substituting in the definition \eqref{jointSI} for
the joint probability, we obtain
\begin{align}
\label{LP}
\mathcal{L}_{t,y_1,...,y_m} {\rm Pr}(\{n_i\},\{y_i\},m)
&=\left\langle \int_0^{\infty} {\rm d}t \, e^{-st} \, I(t_m<t<t_{m+1}) 
\prod_{i=1}^m  \delta_{f_i,n_i} e^{-u_i\tau_i} \,\right\rangle \nonumber \\
&=\left\langle \frac{e^{-st_m}-e^{-st_{m+1}}}{s}
\prod_{i=1}^m  \delta_{f_i,n_i} \, e^{-u_i\tau_i} \right\rangle.
\end{align}
\end{subequations}

We now write $t_m$ and $t_{m+1}$ (the latter defines the end of the phase
$m+1$, which happens after time $t$), as a function of the $\tau_i$'s:
\begin{equation}\label{tm}
t_{m+1}\,=\,t_m+\tau_{m+1}\,= \,\tau_1+...+\tau_{m+1}.
\end{equation}
Here $\tau_{m}$ is the duration of the $m^{\rm th}$ complete phase (which is
longer than the duration of the interrupted phase $\tau^*$).  Using these
relations in \eqref{LP} yields
\begin{align}
\label{lap}
\mathcal{L}_{t,y_1,...,y_m}\, {\rm Pr}(\{n_i\},\{y_i\},m)&=
\left\langle \frac{1-e^{-s \tau_{m+1}}}{s}
\prod_{i=1}^m  \delta_{f_i,n_i}\, e^{-(u_i+s)\tau_i} \right\rangle\,,\nonumber\\
&=
\frac{1-\langle e^{-s \tau} \rangle_1}{s}
\prod_{i=1}^m  \left\langle \delta_{f_i,n_i}\,  e^{-(u_i+s)\tau_i} \right\rangle\,,
\end{align}
where we use the independence of the phase durations in the second line.  We
now denote by $g(n,y)$ the joint probability that the searcher consumes an
amount of food $n$ during this phase and that this phase lasts a time $y$.
We also define the Laplace transform of this joint probability with respect
to $y$ as $\hat{g}(n,s)$.  We then rewrite the ensemble average in
Eq.~\eqref{lap} as a sum over the amount of food consumed in phase $i$ and
over all possible phase durations:
\begin{equation}
\label{rewrite}
\left\langle  \delta_{f_i,n_i}\,e^{-(u_i+s)\tau_i}\right\rangle_1 
= \int_0^{\infty} {\rm d}y'_i \sum_{n_i'=0}^{\infty} \, g(n'_i,y'_i) \, e^{-(u_i+s)y'_i} \, \delta_{n'_i,n_i} =\hat{g}(n_i,s+u_i).
\end{equation}
Substituting Eq.~\eqref{rewrite} into \eqref{lap} gives
\begin{equation}
\label{lap2}
\mathcal{L}_{t,y_1,...,y_m}\, {\rm Pr}(\{n_i\},\{y_i\},m)=\frac{1-\left\langle e^{-s\tau} \right\rangle_1}{s}\prod_{i=1}^m \hat{g}(n_i,s+u_i).
\end{equation} 

We now use the feature that shifting the Laplace variable corresponds to
multiplication by an exponential factor in the time domain
\begin{equation}
\mathcal{L}_t \left( e^{-at} h(t) \right) \equiv \int_0^{\infty} {\rm d}t \;
e^{-st} e^{-at} h(t)=\hat{h}(s+a)\,,
\end{equation}
to obtain the inverse Laplace transform (with respect to the variables $u_i$)
of Eq.~\eqref{lap2}:
\begin{equation}
\label{lapjoint} 
\mathcal{L}_t \,{\rm Pr}(\{n_i\},\{y_i\},m)
=\frac{1-\left\langle e^{-s\tau}\right\rangle_1}{s}
  \prod_{i=1}^m e^{-s y_i} \, g(n_i,y_i).
\end{equation}
We can now express simply the Laplace transform of $F_t$ with respect to $t$
using Eqs.~\eqref{basisSI} and~\eqref{lapjoint}
\begin{align}
\mathcal{L}_t \left\langle e^{-p F_t}\right\rangle
&=\frac{1-\left\langle e^{-s\tau}\right\rangle_1}{s} \sum_{m=0}^{\infty} 
\left( \prod_{i=1}^m\int_0^{\infty} {\rm d}y_i \sum_{n_i=0}^{\infty} e^{-p n_i-s y_i} g(n_i,y_i) \right) \nonumber \\
&=\frac{1-\left\langle e^{-s\tau}\right\rangle_1}{s} \sum_{m=0}^{\infty}  \left(\int_0^{\infty} {\rm d}y \sum_{n=0}^{\infty} e^{-p n-s y} g(n,y) \right)^m .
\end{align}
The expression inside the parentheses is exactly the single-patch ensemble
average $\langle e^{-pf-s \tau} \rangle_1$.  Thus after performing the
geometrical sum over $m$, we obtain Eq.~\eqref{step1}:
\begin{equation}
\label{distribgen}
\mathcal{L}_t \left\langle e^{-pF_t}\right\rangle =
\frac{1-\left\langle e^{-s\tau}\right\rangle_1}
{s\left(1-\langle e^{-pf-s \tau} \rangle_1 \right)}.
\end{equation}

\section{Derivation of Eqs.~\eqref{step2} and \eqref{Q}}
\label{eq5ab}

Having derived the generating function of $F_t$ in the Laplace domain
(Eq.~\eqref{step1}), we now focus on the specific case of one dimension, in
which the searcher uses the give-up criterion that it migrates to another
patch when upon taking $\mathcal{S}$ steps without encountering food.  We
first derive $\langle e^{-pf-s T} \rangle_1$, in the limit of large
$\mathcal{S}$, that we trivially relate to $\langle e^{-pf-s \tau} \rangle_1$
using $\tau=T+Z$. Following~\cite{Benichou:2014}, we treat the time $T$ spent
in a patch, which is normally an (integer) number of steps, as a continuous
variable.  We write the ensemble average $\langle e^{-pf-s T} \rangle_1$ as
\begin{align}
\label{joint}
  \left\langle e^{-pf-sT}\right\rangle_1 =  
\int_0^\infty {\rm d}y  \sum_{n=0}^\infty e^{-pn-sy}\,h(y,n) 
   = \sum_{n=0}^\infty P(n) e^{-pn} \int_0^\infty {\rm d}y \, e^{-sy}\,h(y|n)
\end{align}
where $h(y,n)$ is the joint probability for the searcher to spend a time $y$
in a patch and consume $n$ units of food.  Additionally, $P(n)$ is the
marginal probability that $n$ units of food are consumed in a single patch
when the searcher leaves the patch, integrated over all exploitation
durations, and $h(y|n)$ is the conditional probability to spend a time $y$ in
the patch, given that $n$ units of food have been consumed.

As discussed in Sec.~\ref{formal}, to evaluate $\langle e^{-pf-s T} \rangle_1$ ,
we need to characterize the trajectory of a one-dimensional starving random
walk that has visited $n$ distinct sites at the instant of starvation
(see~\cite{Benichou:2014}).  This trajectory consists of $n-1$ successful
returns to food at either end of a growing food-free interval and a final
lethal excursion.  A return to food is successful if it takes less than
$\mathcal{S}$ steps, so the duration $R_k$ of the $k^{\rm th}$ return
satisfies $R_k<\mathcal{S}$ for $k\in [2,n]$ and
$R_{n+1} \geqslant \mathcal{S}$, corresponding to the walker dying before
reaching one end of the interval.  %The first return in an interval of length
%2 takes no time, because this walker immediately eats food on its initial
%site.  
Each time the walker returns to one end of the interval, it eats the
food there, so that the interval length grows by one lattice spacing to
$(k+1)a$ after $k$ returns (with $a$ the lattice spacing). The lifetime of
this walk is therefore $T=R_2+\ldots+R_n+\mathcal{S}$.  The integral in
Eq.~\eqref{joint} can thus be written as
\begin{align}
\label{Un-SI}
\langle e^{-sT}|n\rangle_1 &\equiv \int_0^\infty {\rm d}y \,    e^{-sy} h(y|n)
\nonumber \\
&= \int_0^\infty {\rm d}y \int_0^\infty {\rm d}r_2 \ldots \int_0^\infty {\rm d}r_n e^{-sy} \delta(y-r_2-\ldots-r_n-\mathcal{S}) \prod_{k=2}^n {\rm Pr}(r_k) \nonumber\\
&=  e^{-s\mathcal{S}}\prod_{k=2}^n \langle e^{-s R_k} \rangle_1 
\equiv  e^{-s\mathcal{S}}\,\, U_n\,.
\end{align}
where Pr$(r_k)$ denotes the probability that the $k^{\rm th}$ return to food
lasts $r_k$ steps. The average $\langle e^{-s R_k} \rangle_1$ is conditioned
on the walker surviving until the $k^{\rm th}$ return and can be expressed as
\begin{equation}
\label{terme1}
\langle e^{-s R_k} \rangle_1 = \frac{\int_0^\mathcal{S}{\rm d} t \, e^{-st} \mathcal{F}_k(t)}
{\int_0^\mathcal{S}{\rm d} t \, \mathcal{F}_k(t)}
 \to \int_0^\mathcal{S}{\rm d} t \,\, e^{-st} \mathcal{F}_k(t)\qquad \mathcal{S}\to\infty\,.
\end{equation}
Here $\mathcal{F}_k(t)$ is the first-passage probability that the walker
first exits an interval of length $ka$ at time $t$ when starting a distance
$a$ from one end.  The denominator is thus the probability that the walker
survives until the $k^{\rm th}$ return, {\it i.e.}, it reaches either end of
a food-free interval of length $ka$ within $\mathcal{S}$ steps.  For large
$\mathcal{S}$, this survival probability equals 1 up to an exponentially
small correction.

To complete the calculation of $\langle e^{-pf-s T} \rangle_1$ , we need to
evaluate the terms $\langle e^{-s R_k} \rangle_1$ that comprise $U_n$ in
Eq.~\eqref{Un-SI}.  In the long-time limit, corresponding to $s\to 0$ and
specifically to $s R_k \ll 1$, each such term in $U_n$ is close to 1.  Thus
it is convenient to first compute $\ln U_n$ and then re-exponentiate:
\begin{align}
\label{Un-SI2}
\ln U_n=\sum_{k=2}^n \ln \langle e^{-s R_k} \rangle 
=\sum_{k=2}^n \ln \left( 1+\langle e^{-s R_k} -1\rangle \right)\,.
\end{align}

We now substitute the explicit expression for the first-passage
probability~\cite{Redners}
\begin{align}
 \mathcal{F}_k(t)=\frac{4\pi D}{\big((k-1)a\big)^2}\,
\sum_{j=0}^\infty & (2j\!+\!1)\, \sin\frac{(2j\!+\!1)\pi}{k-1}  
\exp\left\{-\left[\frac{(2j\!+\!1)\pi}{(k-1)a}\right]^2 Dt\right\}\,,
\end{align}
with the diffusion constant $D=a^2/2$ for a one-dimensional lattice random
walk, in Eq.~\eqref{terme1} to recast \eqref{Un-SI2} as
\begin{align}
\ln U_n=\sum_{k=2}^n \ln \left[ 1+ \frac{4}{ku^2\mathcal{S}}\sum_{j=0}^{\infty}  (2j+1)^2   
\left( \frac{1-e^{-(s\mathcal{S}+(2j+1)^2/u^2)}}
{s+(2j+1)^2/(u^2\mathcal{S})}-\frac{1-e^{-(2j+1)^2/u^2}}{(2j+1)^2/(u^2\mathcal{S})}\right) \right]\,,
\end{align}
where $u\equiv \sqrt{2}\,k/(\pi \sqrt{\mathcal{S}})$.  Since the argument of
the logarithm is close to 1 for $\mathcal{S}\to\infty$, we expand to lowest
order to give
\begin{align}\label{lnUn}
\ln U_n\simeq &\sum_{k=2}^n  \frac{4}{ku^2\mathcal{S}}
\sum_{j=0}^{\infty} (2j+1)^2 
\left( \frac{1-e^{-[s\mathcal{S}+(2j+1)^2/u^2]}}
{s+(2j+1)^2/(u^2\mathcal{S})}-\frac{1-e^{-(2j+1)^2/u^2}}{(2j+1)^2/(u^2\mathcal{S})}\right)\,.
\end{align}
We now introduce $\theta\equiv \sqrt{2}\,n/(\pi \sqrt{\mathcal{S}})$ and take the
continuum limit of Eq.~\eqref{lnUn}, using again $u=\sqrt{2}\,k/(\pi \sqrt{\mathcal{S}})$, to give
\begin{equation}
\ln U(\theta)  \simeq 4 \!\int_0^\theta \!\frac{{\rm d}u}{u}
\sum_{j=0}^{\infty} \left\{\frac{1-e^{-[s\mathcal{S}+{(2j+1)^2}/{u^2}]}}{1+{su^2\mathcal{S}}/{(2j+1)^2}}- \left[  1 - e^{-{(2j+1)^2}/{u^2}}\right]\right\}.
\end{equation}

Furthermore, the distribution $P(n)$ that appears in Eq.~\eqref{joint} was
determined in the continuum limit in Ref.~\cite{Benichou:2014} in terms of
the rescaled variable $\theta$
\begin{equation}
P(\theta)=\frac{4}{\theta}\sum_{j= 0}^{\infty} e^{-{(2j+1)^2}/{\theta^2}}\,
\exp\left\{ -2\sum_{k=0}^{\infty} E_1 \left[{(2k+1)^2}/{\theta^2}\right]\right\},
\end{equation}
where $E_1(x)\equiv \int_1^\infty dt \, e^{-xt}/t$ is the exponential
integral function.  We finally obtain 
\begin{align}
\label{TLdouble}
&\langle e^{-sT-pf}\rangle_1=\int_0^{\infty}  \! \! {\rm d}\theta \, P(\theta)
  \, e^{-p\pi\theta\sqrt{\mathcal{S}/2}-s\mathcal{S}} \,  
\exp\left\{ 4\int_0^\theta \!\frac{{\rm d}u}{u}
\sum_{j=0}^{\infty} \left[\frac{1-e^{-[s\mathcal{S}+(2j+1)^2/u^2]}}
{{1+s\mathcal{S}u^2}/{(2j+1)^2}}-\left(1-e^{-(2j+1)^2/u^2}\right)\!\right]\!\!\right\}\,,
\end{align}
and using the relation $\tau=T+Z$, we obtain Eqs.~\eqref{step2} and
\eqref{Q}.

\section{Derivation of Eqs.~\eqref{mean-var1}}
\label{eq6}

We now extract the moments of $F_t$ in the long-time limit starting from
Eq.~\eqref{distribgen} (which coincides with \eqref{step1}).  These moments
are obtained by expanding the generating function for $p\to 0$:
\begin{equation}
\label{gf-SI}
  \langle e^{-p F_t} \rangle = 1-p\langle F_t \rangle +\frac{p^2}{2} \, \langle F_t^2 \rangle +\ldots\,,
\end{equation}
where $\ldots$ denotes higher-order terms that are negligible as $p\to 0$.
From \eqref{gf-SI}, the temporal Laplace transform is
\begin{equation}
\label{defmom}
\mathcal{L}_t \langle e^{-p F_t} \rangle = \frac{1}{s}-p \, \mathcal{L}_t\langle F_t \rangle +\frac{p^2}{2} \mathcal{L}_t \, \langle F_t^2 \rangle +\dots\,.
\end{equation}

We now straightforwardly expand Eq.~\eqref{step1} in a series in $p$ to give
\begin{equation}\label{devFt}
\mathcal{L}_t \langle e^{-p F_t} \rangle = \frac{1}{s} 
\left[ 1-p \frac{\langle f e^{-s\tau} \rangle_1}{1-\langle e^{-s\tau}
    \rangle_1} +\frac{p^2}{2} \left(\frac{\langle f^2 e^{-s\tau}
      \rangle_1}{1-\langle e^{-s\tau} \rangle_1} +\frac{2 \langle f
      e^{-s\tau} \rangle_1^2}{\left( 1-\langle e^{-s\tau} \rangle_1\right)^2}
  \right)\right]+\ldots \,.
\end{equation}
We identify the first two moments of $F_t$ by comparing Eqs.~\eqref{defmom}
and~\eqref{devFt}, yielding, in the Laplace domain
\begin{subequations}
\begin{align}
& \mathcal{L}_t \langle F_t \rangle =\frac{\langle f e^{-s\tau} \rangle_1}{s \left(1-\langle e^{-s\tau} \rangle_1 \right)} \\
& \mathcal{L}_t \langle F_t^2 \rangle =\frac{1}{s} \left(\frac{\langle f^2 e^{-s\tau} \rangle_1}{1-\langle e^{-s\tau} \rangle_1} +\frac{2 \langle f e^{-s\tau} \rangle_1^2}{\left( 1-\langle e^{-s\tau} \rangle_1\right)^2} \right).
\end{align}
\end{subequations}
The long-time behavior of the moments is given by the small-$s$ expansion of
the above Laplace transforms:
\begin{subequations}
\begin{align}
\mathcal{L}_t \langle F_t \rangle &= \frac{\langle f\rangle_1 }{s^2 \langle
                                    \tau \rangle_1} +\frac{1}{s} \left(
                                    \frac{\langle \tau^2 \rangle_1 \langle f
                                    \rangle_1}{2 \langle \tau
                                    \rangle_1^2}-\frac{\langle f \tau
                                    \rangle_1}{\langle \tau \rangle_1}
                                    \right) +\ldots\,,\\[1em]
\mathcal{L}_t \langle F_t^2 \rangle &=\frac{2 \langle f \rangle_1^2}{s^3
                                      \langle \tau \rangle_1^2}+\frac{1}{s^2}
                                      \left[ \frac{2 \langle f
                                      \rangle_1}{\langle \tau \rangle_1}
                                      \left( \frac{\langle \tau^2 \rangle_1
                                      \langle f \rangle_1}{\langle \tau
                                      \rangle_1^2} - 2 \frac{\langle f\tau
                                      \rangle_1}{\langle \tau \rangle_1}
                                      \right) +\frac{\langle f^2
                                      \rangle_1}{\langle \tau \rangle_1}
                                      \right]+ \ldots \, ,
\end{align}
\end{subequations}
where $\ldots$ indicates lower-order terms in $s$ as $s\to 0$.  Note that two
orders in $s$ are necessary to calculate the variance of $F_t$, as the
leading order terms in $s$ for $\langle F_t^2 \rangle$ and
$\langle F_t \rangle ^2$ cancel.  By performing the inverse Laplace
transform, we finally obtain, in the limit $t\to\infty$,
\begin{subequations}
\begin{align}
\label{moyCt}
\langle F_t \rangle &=\frac{\langle f\rangle_1 }{\langle \tau \rangle_1} t +\left( \frac{\langle \tau^2 \rangle_1 \langle f \rangle_1}{2 \langle \tau \rangle_1^2}-\frac{\langle f \tau \rangle_1}{\langle \tau \rangle_1} \right)+ \ldots\,,\\[1em]
\label{moyCt2}
  \langle F_t^2 \rangle & =\frac{2 \langle f \rangle_1^2}{\langle \tau \rangle_1^2}t^2+ \left[ \frac{2 \langle f \rangle_1}{\langle \tau \rangle_1} \left( \frac{\langle \tau^2 \rangle_1 \langle f \rangle_1}{\langle \tau \rangle_1^2} - 2 \frac{\langle f\tau \rangle_1}{\langle \tau \rangle_1} \right) +\frac{\langle f^2 \rangle_1}{\langle \tau \rangle_1} \right] t+\ldots\,.
\end{align}
\end{subequations}
This yields Eqs.~\eqref{mean-var1}, after using $\tau = T+Z$:
\begin{subequations}
\begin{align}
\langle F_t \rangle &=\frac{\langle f\rangle_1 }{\langle T \rangle_1+Z}\, t +\ldots\label{moyCtdom}\\[1em]
{\rm Var}(F_t) & = \left[ \frac{\langle f \rangle_1^2 {\rm Var}(T)}{(\langle T \rangle_1 +Z)^3}+\frac{{\rm Var}(f)}{\langle T \rangle_1+Z} -\frac{2 \langle f \rangle_1 {\rm Cov}(f,T)}{(\langle T \rangle_1+Z)^2} \right] t +\ldots\,,\label{varCtdom}
\end{align}
\end{subequations}
where the variance Var$(f)$ and covariance Cov$(f,T)$ of $f$ and $T$ were
defined after Eq.~\eqref{mean-var1}.

\section{Derivation of Eqs.~\eqref{mean-var2}}
\label{eq7}

Finally, we obtain $\langle f \rangle_1$, $\langle T \rangle_1$, Var$(T)$ and
$\langle fT \rangle_1$ that appear in Eq.~\eqref{mean-var1} by taking the
small-$p$ and small-$s$ limits of the Laplace transform:
\begin{equation}
\label{devgen}
\langle e^{-sT-pf}\rangle_1 = 1 -s\langle T \rangle_1-p\langle f \rangle_1+
ps \langle Tf \rangle_1 +\frac{s^2}{2} \langle T^2 \rangle_1+\ldots \,,
\end{equation}
where $\ldots$ denotes higher-order terms in $p$ and $s$.  Substituting
Eq.~\eqref{TLdouble} in this expansion gives
\begin{align}
\label{exprexpl}
\langle e^{-sT-pf}\rangle_1 = \Big(1-s\mathcal{S} +\frac{s^2}{2} \mathcal{S}^2 \Big)
\int_0^{\infty} {\rm d}\theta \, P(\theta)  
\Big( 1-p\pi \theta \sqrt{\frac{\mathcal{S}}{2}} +\frac{(p\pi\theta)^2\mathcal{S}}{4} \Big) \left(1+ s\mathcal{S} \mathcal{A}(\theta)+s^2 \mathcal{S}^2\frac{2\mathcal{B}(\theta)+\mathcal{A}^2(\theta)}{2}  \right)+\ldots\,,
\end{align}
with
\begin{align}
\begin{split}
\mathcal{A}(\theta) &\equiv \sum_{j=0}^{\infty} \int_0^\theta {\rm d}u \,\frac{4}{u} \left[ \left( 1+ \frac{u^2}{(2j+1)^2} \right) e^{-{(2j+1)^2}/{u^2}} -\frac{u^2}{(2j+1)^2} \right]\,,\\
\mathcal{B}(\theta) &\equiv \sum_{j=0}^{\infty} \int_0^\theta {\rm d}u
\,\frac{4}{u} \left[ \frac{u^4}{(2j+1)^4}
  -\left(\frac{1}{2}+\frac{u^2}{(2j+1)^2} +\frac{u^4}{(2j+1)^4} \right)
  e^{-{(2j+1)^2}/{u^2}} \right]\,.
\end{split}
\end{align}
Here we have also used the small-$s$ expansion of the expression in the last
exponential in Eq.~\eqref{TLdouble}:
\begin{align}
\frac{1-e^{-[s\mathcal{S}+{(2j+1)^2}/{u^2}]}}{1+{u^2s\mathcal{S}}/{(2j+1)^2}}
-\left(1-e^{-{(2j+1)^2}/{u^2}}\right)
& = s\mathcal{S} \left[ \left( 1+ \frac{u^2}{(2j\!+\!1)^2} \right) e^{-{(2j+1)^2}/{u^2}} -\frac{u^2}{(2j\!+\!1)^2} \right] \nonumber \\
&~~~ + s^2 \mathcal{S}^2 \left[ \frac{u^4}{(2j\!+\!1)^4} -\left(\frac{1}{2}+\frac{u^2}{(2j\!+\!1)^2} +\frac{u^4}{(2j\!+\!1)^4} \right) e^{-{(2j+1)^2}/{u^2}} \right]\ldots \nonumber \,,
\end{align}

From Eq.~\eqref{exprexpl}, the moments of $f$ are simply expressed in terms
of the marginal distribution $P(\theta)$:
\begin{align}
\begin{split}
 \langle f \rangle_1 &=\pi \sqrt{\frac{\mathcal{S}}{2}} \int_0^{\infty} {\rm d}\theta \,\theta \,P(\theta) \equiv K_1 \sqrt{\mathcal{S}}\,, \label{moyN}\\
\langle f^2 \rangle_1 &= \frac{\pi^2 \mathcal{S}}{2} \int_0^{\infty} {\rm d}\theta \,\theta^2\, P(\theta) \,,\\
{\rm Var}(f) &=\langle f^2 \rangle_1 -\langle f \rangle_1^2= \frac{\pi^2 \mathcal{S}}{2} \left[ \int_0^{\infty} \!\!{\rm d}\theta \,\theta^2 \,P(\theta) - \left( \int_0^{\infty} \!\!{\rm d}\theta\, \theta \,P(\theta) \right)^2\, \right] \equiv K_4 \mathcal{S}.
\end{split}
\end{align}
Similarly, identifying~\eqref{exprexpl} with~\eqref{devgen}, we obtain
\begin{align}
\begin{split}
\langle T \rangle_1 &= \left[1-\int_0^{\infty} {\rm d}\theta P(\theta) \mathcal{A}(\theta) \right] \mathcal{S} \equiv K_2 \mathcal{S}\,, \label{moyT}\\
\langle T^2 \rangle_1 &= \left[1+2\int_0^{\infty} {\rm d}\theta P(\theta) \left(\mathcal{B}(\theta)+\frac{1}{2}\mathcal{A}^2(\theta) -\mathcal{A}(\theta)\right) \right] \mathcal{S}^2\,,\\
{\rm Var}(T) &= \left[ \int_0^{\infty} {\rm d}\theta \, P(\theta) \left( 2
    \mathcal{B}(\theta)+ \mathcal{A}^2(\theta) \right)
  -\left(\int_0^{\infty} {\rm d}\theta \,P(\theta) \mathcal{A}(\theta)
  \right)^2 \,\right] \mathcal{S}^2.
\end{split}
\end{align}
Equation~\eqref{devgen} also yields
\begin{align}
\begin{split}
\langle fT \rangle_1 &= \pi \sqrt{\frac{\mathcal{S}^3}{2}} \int_0^{\infty} {\rm d}\theta \,P(\theta) \,\theta \left(1-\mathcal{A}(\theta) \right)\,,\\
{\rm Cov}(f,T) &\equiv\langle fT \rangle_1-\langle f \rangle_1 \langle T
  \rangle_1  = \pi \sqrt{\frac{\mathcal{S}^3}{2}} \int_0^{\infty} \!\!{\rm d}\theta\, P(\theta) \,\theta \left[ \int_0^{\infty} \!\! {\rm d}\varphi \, P(\varphi) \mathcal{A}(\varphi) - \mathcal{A}(\theta) \right].
\end{split}
\end{align}
Finally, we substitute these asymptotic expressions for the moments of $f$
and $T$ into Eq.~\eqref{mean-var1} and obtain the constants $K_1$ to $K_5$
that appear in Eq.~\eqref{mean-var2}:
\begin{align}
\begin{split}
&K_1 \equiv \frac{\pi}{\sqrt{2}} \int_0^{\infty} {\rm d}\theta \, \theta \,P(\theta) \simeq 2.90\ldots\,, \\
&K_2 \equiv 1-\int_0^{\infty} {\rm d}\theta \, P(\theta) \, \mathcal{A}(\theta) \simeq 3.27\ldots\,, \\
&K_3 \equiv \frac{\pi^2}{2} \left[ \int_0^{\infty}  \! \!  {\rm d}\psi\,
\psi P(\psi) \right]^2 \left[ \int_0^{\infty}  \! \! 
{\rm d}\theta \,P(\theta) \left( 2 \mathcal{B}(\theta)+ \mathcal{A}^2(\theta)\right) 
-\left(\int_0^{\infty}  \! \!  {\rm d}\theta \,P(\theta)
  \,\mathcal{A}(\theta) \right)^2 \, \right]  \simeq 16.1\ldots \,,\\
&K_4 \equiv \frac{\pi^2}{2} \left[ \int_0^{\infty} {\rm d}\theta
  \,\theta^2 \,P(\theta) - \left( \int_0^{\infty} {\rm d}\theta \,\theta
    P(\theta) \right)^2 \, \right] \simeq 1.78\ldots\,, \\
&K_5 \equiv  \pi^2 \int_0^{\infty} {\rm d}\psi \,\psi \,
P(\psi)  \int_0^{\infty} {\rm d}\theta \,P(\theta) \,\theta \left[
  \int_0^{\infty} {\rm d}\varphi \,P(\varphi) \mathcal{A}(\varphi) -
  \mathcal{A}(\theta) \right] \simeq 8.51\ldots\,,
\end{split}
\end{align}
where the results are quoted to three-digit accuracy.

\end{widetext}


\begin{thebibliography}{27}
\expandafter\ifx\csname natexlab\endcsname\relax\def\natexlab#1{#1}\fi
\expandafter\ifx\csname bibnamefont\endcsname\relax
  \def\bibnamefont#1{#1}\fi
\expandafter\ifx\csname bibfnamefont\endcsname\relax
  \def\bibfnamefont#1{#1}\fi
\expandafter\ifx\csname citenamefont\endcsname\relax
  \def\citenamefont#1{#1}\fi
\expandafter\ifx\csname url\endcsname\relax
  \def\url#1{\texttt{#1}}\fi
\expandafter\ifx\csname urlprefix\endcsname\relax\def\urlprefix{URL }\fi
\providecommand{\bibinfo}[2]{#2}
\providecommand{\eprint}[2][]{\url{#2}}

\bibitem{Wawerla:2009}
\bibinfo{author}{\bibfnamefont{J.} \bibnamefont{Wawerla}} \bibnamefont{and} \bibinfo{author}{\bibfnamefont{R.~T.} \bibnamefont{Vaughan,}}
\newblock in \emph{\bibinfo{booktitle}{2009 IEEE/RSJ International Conference
  on Intelligent Robots and Systems}}, \bibinfo{pages}{5033--5038}
  (\bibinfo{organization}{IEEE}, \bibinfo{year}{2009}).

\bibitem{Gittins:1973}
\bibinfo{author}{\bibfnamefont{J.~C.} \bibnamefont{Gittins,} }
\newblock \bibinfo{journal}{R\&D Management}
  \textbf{\bibinfo{volume}{3}}, \bibinfo{pages}{71--81} (\bibinfo{year}{1973}).

\bibitem{MP66} R. H. MacArthur and E. R. Pianka, Am.\ Nat.\ \textbf{100}, 603
  (1966).

\bibitem{Pyke:1977}
\bibinfo{author}{\bibfnamefont{G.~H.} \bibnamefont{Pyke,}} \bibinfo{author}{\bibfnamefont{H.~R.} \bibnamefont{Pulliam,}} \bibnamefont{and}
  \bibinfo{author}{\bibfnamefont{E.~L.} \bibnamefont{Charnov,}}
\newblock \bibinfo{journal}{Q. Rev. Biol.}
 \textbf{\bibinfo{volume}{52}}, \bibinfo{pages}{137} (\bibinfo{year}{1977}).

\bibitem{VLRS11} G. M. Viswanathan, M. G. E. da Luz, E. P. Raposo, and
  H. E. Stanley \emph{The Physics of Foraging} (Cambridge University Press,
  Cambridge, UK, 2011).

\bibitem[{\citenamefont{Charnov}(1976)}]{Charnov:1976}
\bibinfo{author}{\bibfnamefont{E.~L.} \bibnamefont{Charnov}},
  \bibinfo{journal}{Theor. Popul. Biol.}
  \textbf{\bibinfo{volume}{9}}, \bibinfo{pages}{129} (\bibinfo{year}{1976}).


\bibitem[{\citenamefont{Stephens and Krebs}(1986)}]{Stephens-DW:1986}
\bibinfo{author}{\bibfnamefont{D.~W.} \bibnamefont{Stephens}} \bibnamefont{and}
  \bibinfo{author}{\bibfnamefont{J.~R.} \bibnamefont{Krebs}},
  \emph{\bibinfo{title}{Foraging Theory}} (\bibinfo{publisher}{Princeton
  University Press, Princeton}, \bibinfo{year}{1986}).

\bibitem[{\citenamefont{B{\'e}nichou et~al.}(2011)\citenamefont{B{\'e}nichou,
  Loverdo, Moreau, and Voituriez}}]{Benichou:2011}
\bibinfo{author}{\bibfnamefont{O.}~\bibnamefont{B{\'e}nichou}},
  \bibinfo{author}{\bibfnamefont{C.}~\bibnamefont{Loverdo}},
  \bibinfo{author}{\bibfnamefont{M.}~\bibnamefont{Moreau}}, \bibnamefont{and}
  \bibinfo{author}{\bibfnamefont{R.}~\bibnamefont{Voituriez}},
  \bibinfo{journal}{Rev. Mod. Phys.} \textbf{\bibinfo{volume}{83}},
  \bibinfo{pages}{81} (\bibinfo{year}{2011}).

\bibitem[{\citenamefont{Oaten}(1977)}]{Oaten:1977}
\bibinfo{author}{\bibfnamefont{A.}~\bibnamefont{Oaten}},
  \bibinfo{journal}{Theor. Popul. Biol.}
  \textbf{\bibinfo{volume}{12}}, \bibinfo{pages}{263} (\bibinfo{year}{1977}).


\bibitem{G80} R. H. Green, Annu.\ Rev.\ Ecol.\ and Systematics \textbf{11},
  1 (1980). %Multivariate Approaches in Ecology: The Assessment of Ecologic
            %Similarity

\bibitem{M82} J. McNamara, Theor.\ Popul.\ Biol.\ \textbf{21}, 269
  (1982). %Optimal patch use in a stochastic environment.

\bibitem{OH88} O. Olsson and N. M. A. Holmgren, Behav.\ Ecol.\ \textbf{9},
  345 (1998).
%The survival-rate-maximizing policy for Bayesian foragers: wait for good news. 

\bibitem{MCS89} E. A. Marschall, P. L. Chesson, and R. A. Steins, Anim.\
  Behav.\ \textbf{37}, 444 (1989).
  % Foraging in a patchy environment: prey-encounter rate and residence time
  % distributions

\bibitem{VB89} T. J. Valone and J. S. Brown, Ecology \textbf{70}, 1800
  (1989).
%Measuring patch assessment abilities of desert granivores.

\bibitem{V91} T. J. Valone, Anim.\ Behav.\ \textbf{41}, 569 (1991).
%Bayesian and prescient assessment: foraging with pre-harvest information. 

\bibitem{OB06} O. Olsson and J. S. Brown, Oikos \textbf{112}, 260 (2006).
%The foraging benefits of information and the penalty of ignorance

\bibitem{V06} T. J. Valone, Oikos \textbf{112}, 252 (2006).
%Are animals capable of Bayesian updating? An empirical review

\bibitem{ECMG07} S. Eliassen, J. Christian, M. Mangel, and J. Giske, Oikos
  \textbf{116}, 513 (2007).
  % Exploration or exploitation: life expectancy changes the value of
  % learning in foraging strategies

\bibitem{PS11} D. J. van der Post and D. Semmann, BMC Evol.\ Biol.\
\textbf{11}, 335 (2011).
%Patch depletion, niche structuring and the evolution of co-operative foraging

\bibitem[{\citenamefont{Iwasa et~al.}(1981)\citenamefont{Iwasa, Higashi, and
  Yamamura}}]{Iwasa:1981}
\bibinfo{author}{\bibfnamefont{Y.}~\bibnamefont{Iwasa}},
  \bibinfo{author}{\bibfnamefont{M.}~\bibnamefont{Higashi}}, \bibnamefont{and}
  \bibinfo{author}{\bibfnamefont{N.}~\bibnamefont{Yamamura}},
  \bibinfo{journal}{Amer. Nat.}  \textbf{\bibinfo{volume}{117}}, \bibinfo{pages}{710}
  (\bibinfo{year}{1981}).

\bibitem[{\citenamefont{Green}(1984)}]{Green:1984}
\bibinfo{author}{\bibfnamefont{R.~F.} \bibnamefont{Green}},
  \bibinfo{journal}{Amer. Nat.} \textbf{\bibinfo{volume}{123}}, \bibinfo{pages}{30}
  (\bibinfo{year}{1984}).

\bibitem[{\citenamefont{Viswanathan et~al.}(1999)\citenamefont{Viswanathan,
  Buldyrev, Havlin, Da~Luz, Raposo, and Stanley}}]{Viswanathan:1999a}
\bibinfo{author}{\bibfnamefont{G.}~\bibnamefont{Viswanathan}},
  \bibinfo{author}{\bibfnamefont{S.~V.} \bibnamefont{Buldyrev}},
  \bibinfo{author}{\bibfnamefont{S.}~\bibnamefont{Havlin}},
  \bibinfo{author}{\bibfnamefont{M.}~\bibnamefont{Da~Luz}},
  \bibinfo{author}{\bibfnamefont{E.}~\bibnamefont{Raposo}}, \bibnamefont{and}
  \bibinfo{author}{\bibfnamefont{H.~E.} \bibnamefont{Stanley}},
  \bibinfo{journal}{Nature} \textbf{\bibinfo{volume}{401}},
  \bibinfo{pages}{911} (\bibinfo{year}{1999}).

\bibitem[{\citenamefont{B\'enichou et~al.}(2005)\citenamefont{B\'enichou,
  Coppey, Moreau, Suet, and Voituriez}}]{Benichou:2005}
\bibinfo{author}{\bibfnamefont{O.}~\bibnamefont{B\'enichou}},
  \bibinfo{author}{\bibfnamefont{M.}~\bibnamefont{Coppey}},
  \bibinfo{author}{\bibfnamefont{M.}~\bibnamefont{Moreau}},
  \bibinfo{author}{\bibfnamefont{P.-H.} \bibnamefont{Suet}}, \bibnamefont{and}
  \bibinfo{author}{\bibfnamefont{R.}~\bibnamefont{Voituriez}},
  \bibinfo{journal}{Phys. Rev. Lett.} \textbf{\bibinfo{volume}{94}},
  \bibinfo{pages}{198101} (\bibinfo{year}{2005}).

\bibitem[{\citenamefont{Oshanin et~al.}(2007)\citenamefont{Oshanin, Wio,
  Lindenberg, and Burlatsky}}]{Oshanin:2007}
\bibinfo{author}{\bibfnamefont{G.}~\bibnamefont{Oshanin}},
  \bibinfo{author}{\bibfnamefont{H.}~\bibnamefont{Wio}},
  \bibinfo{author}{\bibfnamefont{K.}~\bibnamefont{Lindenberg}},
  \bibnamefont{and}
  \bibinfo{author}{\bibfnamefont{S.}~\bibnamefont{Burlatsky}},
  \bibinfo{journal}{J. Phys. Condens. Matter}
  \textbf{\bibinfo{volume}{19}}, \bibinfo{pages}{065142}
  (\bibinfo{year}{2007}).

\bibitem[{\citenamefont{Lomholt et~al.}(2008)\citenamefont{Lomholt, Tal,
  Metzler, and Joseph}}]{Lomholt:2008}
\bibinfo{author}{\bibfnamefont{M.~A.} \bibnamefont{Lomholt}},
  \bibinfo{author}{\bibfnamefont{K.}~\bibnamefont{Tal}},
  \bibinfo{author}{\bibfnamefont{R.}~\bibnamefont{Metzler}}, \bibnamefont{and}
  \bibinfo{author}{\bibfnamefont{K.}~\bibnamefont{Joseph}},
  \bibinfo{journal}{Proc. Natl. Acad. Sci.}
  \textbf{\bibinfo{volume}{105}}, \bibinfo{pages}{11055}
  (\bibinfo{year}{2008}).

\bibitem[{\citenamefont{Bressloff and Newby}(2011)}]{Bressloff:2011}
\bibinfo{author}{\bibfnamefont{P.~C.} \bibnamefont{Bressloff}}
  \bibnamefont{and} \bibinfo{author}{\bibfnamefont{J.~M.} \bibnamefont{Newby}},
  \bibinfo{journal}{Phys. Rev. E} \textbf{\bibinfo{volume}{83}}, \bibinfo{pages}{061139}
  (\bibinfo{year}{2011}).

\bibitem[{\citenamefont{Tejedor et~al.}(2012)\citenamefont{Tejedor, Voituriez,
  and B{\'e}nichou}}]{Tejedor:2012}
\bibinfo{author}{\bibfnamefont{V.}~\bibnamefont{Tejedor}},
  \bibinfo{author}{\bibfnamefont{R.}~\bibnamefont{Voituriez}},
  \bibnamefont{and}
  \bibinfo{author}{\bibfnamefont{O.}~\bibnamefont{B{\'e}nichou}},
  \bibinfo{journal}{Phys. Rev. Lett.} \textbf{\bibinfo{volume}{108}},
  \bibinfo{pages}{088103} (\bibinfo{year}{2012})

\bibitem{WRVL15} M. E. Wosniack, E. P. Raposo, G. M. Viswanathan, and
  M. G. E. da Luz, Phys.\ Rev.\ E {\bf 92}, 062135 (2015).  

\bibitem{NWS15}  B. B. S. Niebuhr, M. E. Wosniack, M. C. Santos, E. P. Raposo,
  G. M. Viswanathan, M. G. E. da Luz, and M. R. Pie, Sci.\ Repts.\ {\bf 5},
  11898 (2015).

\bibitem{RBL03} E. P. Raposo, S. V. Buldyrev, M. G. E. da Luz, M. C. Santos,
  H. E. Stanley, and G. M. Viswanathan, Phys.\ Rev.\ Lett. {\bf 91}, 240601
  (2003).  

\bibitem{McNair:1983} J. N. McNair, Amer.\ Zool.\ {\bf 23}, 303 (1983).

\bibitem[{\citenamefont{Krebs et~al.}(1974)\citenamefont{Krebs, Ryan, and
  Charnov}}]{Krebs:1974}
\bibinfo{author}{\bibfnamefont{J.~R.} \bibnamefont{Krebs}},
  \bibinfo{author}{\bibfnamefont{J.~C.} \bibnamefont{Ryan}}, \bibnamefont{and}
  \bibinfo{author}{\bibfnamefont{E.~L.} \bibnamefont{Charnov}},
  \bibinfo{journal}{Anim. Behav.} \textbf{\bibinfo{volume}{22}},
  \bibinfo{pages}{953} (\bibinfo{year}{1974}).
  
\bibitem[{\citenamefont{Benhamou}(2007)}]{Benhamou:2007}
\bibinfo{author}{\bibfnamefont{S.}~\bibnamefont{Benhamou}},
  \bibinfo{journal}{Ecology} \textbf{\bibinfo{volume}{88}},
  \bibinfo{pages}{1962} (\bibinfo{year}{2007}).

\bibitem[{\citenamefont{Plank and James}(2008)}]{Plank:2008}
\bibinfo{author}{\bibfnamefont{M.}~\bibnamefont{Plank}} \bibnamefont{and}
  \bibinfo{author}{\bibfnamefont{A.}~\bibnamefont{James}},
  \bibinfo{journal}{J. R. Soc. Interface}
  \textbf{\bibinfo{volume}{5}}, \bibinfo{pages}{1077} (\bibinfo{year}{2008}).

\bibitem[{\citenamefont{Nolting et~al.}(2015)\citenamefont{Nolting, Hinkelman,
  Brassil, and Tenhumberg}}]{Nolting:2015}
\bibinfo{author}{\bibfnamefont{B.~C.} \bibnamefont{Nolting}},
  \bibinfo{author}{\bibfnamefont{T.~M.} \bibnamefont{Hinkelman}},
  \bibinfo{author}{\bibfnamefont{C.~E.} \bibnamefont{Brassil}},
  \bibnamefont{and}
  \bibinfo{author}{\bibfnamefont{B.}~\bibnamefont{Tenhumberg}},
  \bibinfo{journal}{Ecol. Complexity} \textbf{\bibinfo{volume}{22}},
  \bibinfo{pages}{126} (\bibinfo{year}{2015}).
  
\bibitem[{\citenamefont{Gueudr{\'e} et~al.}(2014)\citenamefont{Gueudr{\'e},
  Dobrinevski, and Bouchaud}}]{Gueudre:2014}
\bibinfo{author}{\bibfnamefont{T.}~\bibnamefont{Gueudr{\'e}}},
  \bibinfo{author}{\bibfnamefont{A.}~\bibnamefont{Dobrinevski}},
  \bibnamefont{and} \bibinfo{author}{\bibfnamefont{J.-P.}
  \bibnamefont{Bouchaud}}, \bibinfo{journal}{Phys. Rev. Lett.}
  \textbf{\bibinfo{volume}{112}}, \bibinfo{pages}{050602}
  (\bibinfo{year}{2014}).

\bibitem{March:1991}
\bibinfo{author}{\bibfnamefont{ J.~G.} \bibnamefont{March}}
\newblock \bibinfo{journal}{Organ. Sci.}
  \textbf{\bibinfo{volume}{2}}, \bibinfo{pages}{71} (\bibinfo{year}{1991}).


  \bibitem[{\citenamefont{Cohen et~al.}(2007)\citenamefont{Cohen, McClure, and
  Yu}}]{Cohen:2007}
\bibinfo{author}{\bibfnamefont{J.~D.} \bibnamefont{Cohen}},
  \bibinfo{author}{\bibfnamefont{S.~M.} \bibnamefont{McClure}},
  \bibnamefont{and} \bibinfo{author}{\bibfnamefont{A.~J.} \bibnamefont{Yu}},
  \bibinfo{journal}{Phil. Trans. R. Soc. B} \textbf{\bibinfo{volume}{362}}, \bibinfo{pages}{933}
  (\bibinfo{year}{2007}).

\bibitem[{\citenamefont{Weiss}(1994)}]{Weiss}
\bibinfo{author}{\bibfnamefont{G.~H.} \bibnamefont{Weiss}},
  \emph{\bibinfo{title}{Aspects and Applications of the Random Walk}} (\bibinfo{publisher}{North-Holland}, \bibinfo{year}{1994}).

\bibitem[{\citenamefont{Hughes}(1996)}]{Hughes}
\bibinfo{author}{\bibfnamefont{B.~D.} \bibnamefont{Hughes}},
  \emph{\bibinfo{title}{Random Walks and Random Environments}} (\bibinfo{publisher}{Clarendon Press Oxford}, \bibinfo{year}{1996}).

\bibitem[{\citenamefont{Godreche and Luck}(2001)}]{Godreche:2001}
\bibinfo{author}{\bibfnamefont{C.}~\bibnamefont{Godreche}} \bibnamefont{and}
  \bibinfo{author}{\bibfnamefont{J.~M.}~\bibnamefont{Luck}},
  \bibinfo{journal}{J. Stat. Phys.}
  \textbf{\bibinfo{volume}{104}}, \bibinfo{pages}{489} (\bibinfo{year}{2001}).

\bibitem[{\citenamefont{B\'enichou and Redner}(2014)}]{Benichou:2014}
\bibinfo{author}{\bibfnamefont{O.}~\bibnamefont{B\'enichou}} \bibnamefont{and}
  \bibinfo{author}{\bibfnamefont{S.}~\bibnamefont{Redner}},
  \bibinfo{journal}{Phys. Rev. Lett.} \textbf{\bibinfo{volume}{113}},
  \bibinfo{pages}{238101} (\bibinfo{year}{2014}).
  
\bibitem{CBR16} M. Chupeau, O. B\'enichou, S. Redner, J. Phys.\ A: Math.\
  Theor.\ {\bf 49}, 394003 (2016).

\bibitem[{\citenamefont{Yuste et~al.}(2013)\citenamefont{Yuste, Abad, and
  Lindenberg}}]{Yuste:2013}
\bibinfo{author}{\bibfnamefont{S.~B.} \bibnamefont{Yuste}},
  \bibinfo{author}{\bibfnamefont{E.}~\bibnamefont{Abad}}, \bibnamefont{and}
  \bibinfo{author}{\bibfnamefont{K.}~\bibnamefont{Lindenberg}},
  \bibinfo{journal}{Phys. Rev. Lett.} \textbf{\bibinfo{volume}{110}},
  \bibinfo{pages}{220603} (\bibinfo{year}{2013}).


\bibitem{R10} A. M. Reynolds, Physica A {\bf 389}, 4740 (2010).

\bibitem{H79} B. Heinrich, Oecologia, {\bf 40}, 235 (1979).

\bibitem{R12} A. M. Reynolds, J. R. Soc.\ Interface {\bf 9}, 1568 (2012).

\bibitem{BCRLMC16} F. Bartumeus, D. Campos, W. S. Ryu, R. Lloret-Cabot,
  V. M\'endez, and J. Catalan, Ecol.\ Lett.\ {\bf 19}, 1299 (2016).

\bibitem{RB09} A. M. Reynolds and F. Bartumeus, J. Theor.\ Biol.\ {\bf 260},
  98 (2009).

\bibitem[{\citenamefont{Evans and Majumdar}(2011{\natexlab{a}})}]{Evans:2011a}
\bibinfo{author}{\bibfnamefont{M.~R.} \bibnamefont{Evans}} \bibnamefont{and}
  \bibinfo{author}{\bibfnamefont{S.~N.} \bibnamefont{Majumdar}},
  \bibinfo{journal}{Phys. Rev. Lett.} \textbf{\bibinfo{volume}{106}},
  \bibinfo{pages}{160601} (\bibinfo{year}{2011}{\natexlab{a}}).

\bibitem[{\citenamefont{Evans and Majumdar}(2011{\natexlab{b}})}]{Evans:2011b}
\bibinfo{author}{\bibfnamefont{M.~R.} \bibnamefont{Evans}} \bibnamefont{and}
  \bibinfo{author}{\bibfnamefont{S.~N.} \bibnamefont{Majumdar}},
  \bibinfo{journal}{J. Phys. A: Math. \& Theor.}
  \textbf{\bibinfo{volume}{44}}, \bibinfo{pages}{435001}
  (\bibinfo{year}{2011}{\natexlab{b}}).

\bibitem[{\citenamefont{Pal et~al.}(2015)\citenamefont{Pal, Kundu, and
  Evans}}]{Pal:2015}
\bibinfo{author}{\bibfnamefont{A.}~\bibnamefont{Pal}},
  \bibinfo{author}{\bibfnamefont{A.}~\bibnamefont{Kundu}}, \bibnamefont{and}
  \bibinfo{author}{\bibfnamefont{M.~R.} \bibnamefont{Evans}},
  \bibinfo{journal}{J. Phys. A: Math. \& Theor.}
  \textbf{\bibinfo{volume}{49}}, \bibinfo{pages}{225001}
  (\bibinfo{year}{2016}).

\bibitem[{\citenamefont{Nagar and Gupta}(2015)}]{Nagar:2015}
\bibinfo{author}{\bibfnamefont{A.}~\bibnamefont{Nagar}} \bibnamefont{and}
  \bibinfo{author}{\bibfnamefont{S.}~\bibnamefont{Gupta}},
  \bibinfo{journal}{arXiv preprint arXiv:1512.02092}  (\bibinfo{year}{2015}).

\bibitem[{\citenamefont{Eule and Metzger}(2016)}]{Eule:2016}
\bibinfo{author}{\bibfnamefont{S.}~\bibnamefont{Eule}} \bibnamefont{and}
  \bibinfo{author}{\bibfnamefont{J.~J.} \bibnamefont{Metzger}},
  \bibinfo{journal}{New J. Phys.} \textbf{\bibinfo{volume}{18}},
  \bibinfo{pages}{033006} (\bibinfo{year}{2016}).


\bibitem{G01a} C. Godreche and J. M. Luck, J. Stat.\ Phys.\ {\bf 104}, 489
  (2001).

\bibitem{Redners} S. Redner, {\it A Guide to First-Passage Processes},
  (Cambridge University Press, Cambridge, UK, 2001).

\end{thebibliography}
\end{document}